\documentstyle[pre,aps]{revtex}
\begin{document}
\draft
\preprint{}
\title{Equivalence  of two approaches for the inhomogeneous density
in the canonical ensemble}
\author{J.A.\ White and S. Velasco}
\address{ Departamento de F\'{\i}sica Aplicada,  Facultad de Ciencias,
Universidad de Salamanca, 37008 Salamanca, Spain}

\date{\today}
\maketitle
\begin{abstract} In this article we show that the inhomogeneous
density obtained from a  density-functional theory of classical
fluids in the canonical ensemble (CE), recently presented by White
{\it et al} [Phys. Rev. Lett. {\bf 84}, 1220 (2000)], is equivalent to
first order to the result of the series expansion of the CE
inhomogeneous density introduced by Gonz\'alez {\it et al} [Phys. Rev.
Lett. {\bf 79}, 2466 (1997)].
\end{abstract}

\pacs{PACS number(s) 61.20.Gy, 68.45.-v}

A statistical mechanics ensemble is a collection of identical
systems under the same external conditions. Although the choice of a
particular ensemble for studying a concrete system should be guided
by the conditions in which the system is found, one can choose
---due to mathematical or computational convenience--- any
ensemble for analyzing the equilibrium properties of the system.
This way of proceeding,  based on the equivalence of the ensembles
in the thermodynamic limit,  is only  justified for  systems with a
very large number of particles. For small systems, however, the
ensembles are no longer equivalent and  the external conditions
must determine the choice of ensemble.

In this context,  the use of density-functional theory (DFT) for the
study of classical  inhomogeneous fluids has been usually limited to
the grand canonical ensemble  (GCE), where the temperature $T$ and
the chemical potential $\mu$ are fixed by an external reservoir.  A
large variety of inhomogeneous situations has been successfully
studied by means of DFT  in the GCE
\cite{Evans89,Evans92,Lowen94}. These situations include fluids
confined in narrow pores or capillaries\cite{Evans90}, or even
spherical cavities \cite{Calleja91,Kim96,Gonzalez98}, which are
implicitly assumed to be open, i.e., allowing exchange of particles
with a reservoir.  This assumption is crucial for situations with a
small number of particles where, depending on the choice of
ensemble,  important differences may arise in the equilibrium
microscopic structure of the system
\cite{Gonzalez98,Gonzalez97}.  If one wishes to investigate the
properties of a small {\it closed} system at temperature $T$, the
study must be  performed in such a way that one obtains results in
the canonical ensemble (CE) because the  number of particles $N$ is
fixed. In DFT this goal can be achieved by means of two different
approaches. On one hand the DFT could be formulated in the canonical
ensemble
\cite{Parr}, with a minimum free-energy principle with fixed $T$
and $N$, and an appropriate CE functional. Very recently, this
approach has been explicitly realized \cite{White00} by considering
an approximate expression for the CE functional. On the other hand,
one can perform a conventional DFT study in the GCE and then relate
the obtained properties to those of the CE. This approach was
followed in Refs.
\cite{Gonzalez98,Gonzalez97} where the CE density profile  of a
hard-sphere fluid in a  small spherical cavity was calculated by
means of a series expansion in terms of the corresponding GCE
profile. The aim of the present paper is to show that these two
approaches yield equivalent results to order
$1/N$. For clarity we start with a brief summary of the main results
of the two approaches.

The first approach is based on the following series expansion of the
CE density profile
$\rho_{\text{c}}({\bf r})$ in terms of  its corresponding GCE density
profile
$\rho_{\text{gc}}({\bf r})$:
\begin{equation}
\rho_{\text{c}}({\bf r})=\rho_{\text{gc}}({\bf r})-
\frac{1}{2}
\Delta^2(N)
\frac{\partial^2}{\partial{\langle N \rangle}^2}
\rho_{\text{gc}}({\bf r}) + O(\frac{1}{{\langle N \rangle}^2})
\label{a1}
\end{equation} where the grand canonical profile is obtained for a
chemical potential $\mu$ that leads to an average number of
particles $\langle N
\rangle$  equal to $N$ ---the fixed integer number of particles in the
CE, and
$\Delta^2(N)\equiv \langle N^2 \rangle- \langle N \rangle^2$ is the
mean square fluctuation of the number of particles in the GCE.
Higher-order terms  in the above expansion  also depend on
fluctuations and on variations of the GCE profile w.r.t.
$\langle N \rangle$ \cite{Gonzalez98,Gonzalez97}.  In  DFT, the grand
canonical profile $\rho_{\text{gc}}$ is the solution of the usual GCE
Euler-Lagrange equation  with chemical potential $\mu$ and external
potential $V_{\text{ext}}({\bf r})$,
\begin{equation}
\label{a2a} {\delta  {\cal F}_{\text{gc}}[\rho] \over \delta \rho({\bf
r})}\Biggr|_{\rho=\rho_{\text{gc}}}+ V_{\text{ext}}({\bf r}) =\mu,
\end{equation} where ${\cal F}_{\text{gc}}[\rho]$ is the GCE
free-energy functional. For a chemical potential leading to a given
$\langle N \rangle$, Eq.~(\ref{a2a}) can be rewritten
as~\cite{Gonzalez98}:
\begin{equation}
\rho_{\text{gc}}({\bf r})=  \langle N \rangle \exp \left[{-\beta
V_{\text{ext}}({\bf r})+ c^{(1)}({\bf r};[\rho_{\text{gc}}])}\right]
\bigg/ \int  d{\bf r} \exp \left[{-\beta V_{\text{ext}}({\bf r})+
c^{(1)}({\bf r};[\rho_{\text{gc}}])}\right],
\label{a2}
\end{equation} where  $\beta=1/k_{\text{B}} T$ is the inverse
temperature and $c^{(1)}$ is the one-body direct correlation function
\begin{equation} c^{(1)}({\bf r};[\rho])=-\beta{{\delta (   {\cal
F}_{\text{gc}}[\rho] -{\cal F}_{\text{gc-id}}[\rho]}  )  \over
{\delta\rho({\bf r})}} \, ,
\label{a3}
\end{equation} being ${\cal F}_{\text{gc-id}}$ the usual ideal-gas
free-energy. This correlation function is the first member of the
direct correlation hierarchy
\begin{equation} c^{(n)}({\bf r}_{1},...,{\bf
r}_{n};[\rho])=-\beta{{\delta^{n} (   {\cal F}_{\text{gc}}[\rho] -{\cal
F}_{\text{gc-id}}[\rho]}  )
\over {\delta\rho({\bf r}_{1})\cdots\delta\rho({\bf r}_{n})}} \, .
\label{a4}
\end{equation} From Eq. (\ref{a2}), in DFT it is possible to obtain
density profiles normalized for a given $\langle N \rangle$. This
allows for obtaining approximate CE profiles using (\ref{a1}) where
the derivatives w.r.t. $\langle N \rangle$ are calculated numerically.
(We note that, using the thermodynamic identity
$\Delta^2(N)=\partial \langle N \rangle /\partial (\beta\mu)$, the
mean square fluctuation can also be expressed as a derivative w.r.t.
$\langle N
\rangle$.) This procedure was used in \cite{Gonzalez98,Gonzalez97}
to obtain the CE density profile of a hard-sphere fluid confined in a
hard spherical cavity.

The second approach consists of an approximate expression for the
free energy functional in the CE. On the basis of the standard saddle
point relation  between  the CE Helmholtz free energy and the GCE
grand potential~\cite{Zubarev} the following approximation for the
CE free-energy functional $\cal F_{\text{c}}$ was proposed in
Ref.~\cite{White00}:
\begin{equation}
\beta {\cal F}_{\text{c}}[\rho]
\approx  \beta {\cal F}_{\text{gc}}[\rho] +\frac{1}{2} \log {{2
\pi}\Delta^2(N;[\rho])}
\label{b1}
\end{equation}  where the functional dependence of the GCE mean
square fluctuation $\Delta^2(N)$ is made explicit. Since we are now
working in the canonical ensemble, the equilibrium density profile
$\rho_{\text{c}}({\bf r})$ is obtained by minimizing the functional
${\cal F}_{\text{c}}[\rho] +\int  d{\bf r}\rho({\bf
r})V_{\text{ext}}({\bf r})$ subject to the constraint
\begin{equation}
\label{b1b}
\int d{\bf r} \rho_{\text{c}}({\bf r})=N.
\end{equation} Using the Lagrange multiplier technique one obtains
\cite{White00}
\begin{equation}
\label{b1c} {\delta  {\cal F}_{\text{c}}[\rho] \over \delta \rho({\bf
r})}\Biggr|_{\rho=\rho_{\text{c}}}+ V_{\text{ext}}({\bf r}) =\lambda
\end{equation} where the Lagrange multiplier $\lambda$ must be
calculated from the constraint~(\ref{b1b}). This equation can be
re-expressed as
\begin{eqnarray}
\rho_{\text{c}}({\bf r})&=&  {N   \exp \Bigl[{-\beta
V_{\text{ext}}({\bf r}) +c^{(1)}({\bf r};[\rho_{\text{c}}]) + \xi ({\bf
r};[\rho_{\text{c}}]) }\Bigr]
 \Bigg/  \int d{\bf r}} \nonumber \\
 &&\times
 \exp \Bigl[{-\beta V_{\text{ext}}({\bf r}) +c^{(1)}({\bf
r};[\rho_{\text{c}}]) +
\xi ({\bf r};[\rho_{\text{c}}]) }\Bigr]
\label{b2}
\end{eqnarray} where
\begin{equation}
\xi ({\bf r}; [\rho]) \equiv -\beta{ \delta (  {\cal F}_{\text{c}}[\rho] -
{\cal F}_{\text{gc}}[\rho] )
\over \delta\rho({\bf r}) }\,,
\label{b3}
\end{equation} which, for the saddle-point (SP) approximation
(\ref{b1}), becomes
\begin{equation}
\xi ({\bf r};[\rho]) \approx \xi_{\text{sp}} ({\bf r};[\rho]) =
-\frac{1}{2}
\frac{1}{ \Delta^2(N;[\rho]) } { \delta \Delta^2(N;[\rho]) \over
\delta\rho({\bf r})} .
\label{b4}
\end{equation} In order to calculate $\xi_{\text{sp}} ({\bf r};[\rho])$
it is important to express the mean square fluctuation
$\Delta^2(N;[\rho])$ as a functional of the density. This can be  done
conveniently in the GCE by  means of the density-density correlation
function
\cite{Evans92,Henderson92}
\begin{equation}
\label{b5}
 {\cal G}({\bf r}_1,{\bf r}_2)=\beta^{-1}{\delta  \rho({\bf r}_{1})
\over
\delta (\mu-V_{\text{ext}}({\bf r}_{2}))},
\end{equation} since this function normalizes to the mean square
fluctuation, i.e.,
\begin{equation}
\Delta^2(N)= \int\int d{\bf r}_{1} d{\bf r}_{2} {\cal G}({\bf r}_1,{\bf
r}_2)  \, .
\label{b6}
\end{equation} In addition, taking into account that  ${\cal G}$ is the
functional inverse of the second derivative of the GCE free-energy
\begin{eqnarray}
\label{b7}
 {\cal G}^{-1}({\bf r}_1,{\bf r}_2)&=&\beta {\delta
(\mu-V_{\text{ext}}({\bf r}_{1})) \over
\delta \rho({\bf r}_{2}) } =\beta {\delta^2  {\cal F}_{\text{gc}}[\rho]
\over
\delta \rho({\bf r}_{1})\rho({\bf r}_{2})} \nonumber \\ &=& {1 \over
\rho({\bf r}_{1}) } \delta({\bf r}_{1}-{\bf r}_{2}) -c^{(2)}({\bf
r}_{1},{\bf r}_{2}) \,,
\end{eqnarray} and satisfies the Ornstein-Zernike relation
\cite{Evans92,Henderson92}
\begin{equation}
\int d{\bf r}_{2} {\cal G}^{-1}({\bf r}_1,{\bf r}_2) {\cal G}({\bf
r}_2,{\bf r}_3) =
 \delta({\bf r}_{1}-{\bf r}_{3})  \, ,
\label{b8}
\end{equation} one obtains \cite{White00}
\begin{eqnarray}
\Delta^2(N;[\rho])&=&\int d{\bf r} \Gamma({\bf r})  \label{b9a} \\
\frac{\delta}{\delta \rho ({\bf r}) }\Delta^2(N;[\rho]) &=&
\int\int  d{\bf r}_{1}  d{\bf r}_{2}  {\delta {\cal G}^{-1}({\bf r},{\bf
r}_{1})
\over \delta \rho({\bf r}_{2}) }
\Gamma({\bf r}_{1}) \Gamma({\bf r}_{2})\label{b9b}
\\
 &=&
\biggl( {\Gamma({\bf r}) \over \rho({\bf r})}
\biggr)^2 +
\int\int d{\bf r}_{1} d{\bf r}_{2} c^{(3)}({\bf r},{\bf r}_{1},{\bf r}_{2})
 \Gamma({\bf r}_1)
 \Gamma({\bf r}_2)  \,
\label{b9c}
\end{eqnarray} where
\begin{equation}
\Gamma({\bf r})\equiv  \int d{\bf r}_{1} {\cal G}({\bf r},{\bf r}_{1})
\, ,
\label{b10}
\end{equation} is obtained from the following averaged
Ornstein-Zernike relation
\begin{equation}
\Gamma({\bf r}) =\rho({\bf r}) + \rho ({\bf r})
  \int  d{\bf r}_{1} \Gamma({\bf r}_{1}) c^{(2)}({\bf r},{\bf r}_{1})\, .
\label{b11}
\end{equation} In deriving (\ref{b9b}) we have considered the
functional derivative w.r.t. density of the Ornstein-Zernike relation
(\ref{b8}) and exploited the fact that ${\cal G}$ and ${\cal G}^{-1}$
are functional inverses. We note that the key difference between the
GCE result (\ref{a2}) and the CE density  (\ref{b2}) is the term $\xi
({\bf r};[\rho_{\text{c}}])$. We also note that, in this approach, using
(\ref{b2}) one directly obtains the CE profile while, in the previous
approach, the result (\ref{a2}) of GCE-DFT had to be inserted into
(\ref{a1}) in order to obtain an approximation for the CE equilibrium
density. In what follows we shall show that both approaches, agree
to first order, though they yield different results due to higher-order
contributions in the saddle point approach.  We first derive some
useful relations and then we show the equivalence to first order of
the approaches.

Our starting point is the well-known result of GCE density functional
theory that, for given intermolecular potential, chemical potential
$\mu$, and temperature  $T$, only one external potential can
determine a specified equilibrium density profile
\cite{Evans89,Evans92}. Thus there must exist an external potential
$\bar V_{\text{ext}}({\bf r})$ so that its corresponding GCE
equilibrium density
$\rho_{\text{gc}}({\bf r};[\bar V_{\text{ext}}])$   (here the functional
dependence of
$\rho_{\text{gc}}$ is made explicit) is equal to the CE result
$\rho_{\text{c}}$.  Performing a functional expansion of
$\rho_{\text{gc}}({\bf r};[\bar V_{\text{ext}}])$  about
$V_{\text{ext}}$ and using definition (\ref{b5}) we obtain:
\begin{eqnarray}
\label{e2}
\beta^{-1}\Delta \rho({\bf r})&=&
\int  d{\bf r}_{1}  {\cal G}({\bf r},{\bf r}_1)
 \Delta V_{\text{ext}}({\bf r}_{1})\nonumber\\ &&+{1\over 2!}
\int  d{\bf r}_{1} d{\bf r}_{2}  {\delta  {\cal G}({\bf r},{\bf r}_1)
\over
\delta (\mu-V_{\text{ext}}({\bf r}_{2}))}
\Delta V_{\text{ext}}({\bf r}_{1})
\Delta V_{\text{ext}}({\bf r}_{2})+\dots,
\end{eqnarray} where
\begin{equation}
\label{ec2a}
\Delta \rho({\bf r}) =\rho_{\text{gc}}({\bf r};[\bar
V_{\text{ext}}])-\rho_{\text{gc}}({\bf r};[V_{\text{ext}}])=
\rho_{\text{c}}({\bf r})-\rho_{\text{gc}}({\bf r})
\end{equation} and
\begin{equation}
\label{ec2b}
\Delta V_{\text{ext}}({\bf r}) =V_{\text{ext}}({\bf r})-\bar
V_{\text{ext}}({\bf r})\, .
\end{equation} Therefore,  Eq. (\ref{e2}) provides a link  between
$\Delta
\rho$ and
$\Delta V_{\text{ext}}$ via a functional expansion where the
coefficients belong to the standard distribution function hierarchy.

At this point we would like to emphasize the  role played by $\bar
V_{\text{ext}}$  in the present work.  $\bar V_{\text{ext}}$ is the
external potential that, at chemical potential $\mu$, yields the
canonical profile
$\rho_{\text{c}}$  {\it in a GCE approach} and this implies that
$\rho_{\text{c}}$ is the  solution of the following GCE Euler-Lagrange
equation
\begin{equation}
\label{e5} {\delta  {\cal F}_{\text{gc}}[\rho] \over \delta \rho({\bf
r})}\Biggr|_{\rho=\rho_{\text{c}}}+
\bar V_{\text{ext}}({\bf r}) =\mu \, .
\end{equation} This fact makes meaningful the use of functionals of
$\rho_{\text{c}}$ like
${\cal G}^{-1}({\bf r}_1,{\bf r}_2;[\rho_{\text{c}}])=$
 $\beta {\delta  (\mu-{\bar V}_{\text{ext}}({\bf r}_{1})) /
\delta \rho_{\text{c}}({\bf r}_{2}) }$, its inverse,
${\cal G}({\bf r}_1,{\bf r}_2;[\rho_{\text{c}}])=$ ${\cal G}({\bf
r}_1,{\bf r}_2;[{\bar V}_{\text{ext}}[\rho_{\text{c}}]])=$
$\beta^{-1} { \delta \rho_{\text{c}}({\bf r}_{1})  /
\delta  (\mu-{\bar V}_{\text{ext}}({\bf r}_{2}))}$, or the mean square
fluctuation
$\Delta^2(N;[\rho_{\text{c}}])$. On the other hand, since
$\rho_{\text{c}}({\bf r})$ is the equilibrium density profile in the
canonical ensemble, it is the solution of the CE Euler-Lagrange
equation (\ref{b1c}) where, in comparison with this GCE equation,
the free energy is ${\cal F}_{\text{c}}$ ---the CE functional, the
external potential is
$V_{\text{ext}}$, and  the Lagrange multiplier
$\lambda$ is used in the place of the chemical potential  $\mu$.
From Eqs. (\ref{b1c}) and  (\ref{e5}) and definition (\ref{b3}) we
obtain
\begin{equation}
\label{e5bisbis}
\Delta V_{\text{ext}}({\bf r})=\lambda-\mu+\beta^{-1}  \xi ({\bf
r};[\rho_{\text{c}}]).
\end{equation} In the uniform limit, where $\rho_{\text{gc}}({\bf r})
\to
\rho_{0}\equiv\langle N \rangle/V =N/V$ and also
$\rho_{\text{c}}({\bf r})
\to \rho_{0}$, from  (\ref{a2a})  and (\ref{e5})   one has $\Delta
V_{\text{ext}} ({\bf r}) = 0$ and thus
\begin{equation}
\label{e5b}
\mu-\lambda=\beta^{-1}\xi (\rho_{0}),
\end{equation} and  (\ref{e5bisbis}) can be rewritten as
\begin{equation}
\label{e5bisbisbis}
\beta \Delta V_{\text{ext}}({\bf r})=\xi ({\bf r};[\rho_{\text{c}}])-\xi
(\rho_{0}).
\end{equation} Using the saddle-point approximation
$\xi_{\text{sp}}$ [Eq. (\ref{b4})], this expression could be employed
in (\ref{e2}) to obtain an approximation for the difference $\Delta
\rho$. Conversely, since Eq. (\ref{a1}) gives an approximation for
$\Delta \rho$, an expansion inverse to (\ref{e2}) would provide a way
to obtain $\xi$. This inverse expression can be easily derived by
substituting the functional expansion of ${\delta  {\cal
F}_{\text{gc}}[\rho] / \delta
\rho({\bf r})}$ about
$\rho_{\text{gc}}$  in Eq. (\ref{e5}). Using definition (\ref{b7}), we
obtain
\begin{eqnarray}
\label{e7}
\beta\Delta V_{\text{ext}}({\bf r}) &=&
\int  d{\bf r}_{1}  {\cal G}^{-1}({\bf r},{\bf r}_{1}) \Delta \rho ({\bf
r}_{1})\nonumber\\ &&+{1\over 2!}
\int  d{\bf r}_{1} d{\bf r}_{2}  {\delta  {\cal G}^{-1}({\bf r},{\bf
r}_{1}) \over
\delta \rho_{\text{gc}}({\bf r}_{2})}
 \Delta \rho ({\bf r}_{1})\Delta \rho ({\bf r}_{2}) +\dots
\end{eqnarray}   where we have exploited the fact that
$\rho_{\text{gc}}({\bf r})$ is the solution of the usual GCE
Euler-Lagrange equation  (\ref{a2a}).

Expansions (\ref{e2}) and (\ref{e7}) are asymptotically  exact
relations linking $\Delta \rho$ and $\Delta V_{\text{ext}}$. However,
these expansions need to be truncated in order to become suitable for
practical applications. In particular, to first order, Eq. (\ref{e2})
becomes
\begin{equation}
\label{e10}
\beta^{-1}\Delta \rho({\bf r}) \approx
\int  d{\bf r}_{1}  {\cal G}({\bf r},{\bf r}_1)
 \Delta V_{\text{ext}}({\bf r}_{1}),
\end{equation} and  Eq. (\ref{e7}) reduces to
\begin{equation}
\label{e11}
\beta\Delta V_{\text{ext}}({\bf r}) \approx
\int  d{\bf r}_{1}  {\cal G}^{-1}({\bf r},{\bf r}_{1}) \Delta \rho ({\bf
r}_{1}).
\end{equation}   Approximations (\ref{e10}) and (\ref{e11}) are, by
virtue of Eq. (\ref{b8}), equivalent equations; this fact shows the
consistency of the truncation of the expansions.  Either (\ref{e10})
or (\ref{e11}) provide a simple (first order) relation  between the
differences $\Delta\rho$ and
$\Delta V_{\text{ext}}$. Using these equations,  we shall show that
the approximation (\ref{a1}) for $\Delta\rho$ is equivalent to first
order to the saddle point approximation $\xi_{\text{sp}}$ [Eq.
(\ref{b4})].  By considering the derivative of the GCE Euler-Lagrange
equation (\ref{a2a}) w.r.t.
$\langle N \rangle$ at fixed $V_{\text{ext}}$, one obtains the exact
relation
\begin{equation}
\label{e13}
\int  d{\bf r}_1\, {\cal G}^{-1}({\bf r},{\bf r}_1) {\partial
\rho_{\text{gc}}({\bf r}_1)\over \partial \langle N \rangle} = {1 \over
\Delta^{2}(N)}
\end{equation}  where  the chain rule for functional differentiation
together with  definition (\ref{b7}) and the identity $\partial \langle
N \rangle /
\partial (\beta \mu)=\Delta^{2}(N)$ have been used.  This equation
can be rewritten, via Eqs. (\ref{b8}) and (\ref{b10}), as
\begin{equation}
\label{e14} {\partial \rho_{\text{gc}}({\bf r})\over \partial \langle N
\rangle} = {\Gamma({\bf r}) \over \Delta^{2}(N)} .
\end{equation}  Differentiating this equation w.r.t. $\langle N
\rangle$ and using  Eq. (\ref{a1}),  we obtain
\begin{equation}
\label{e16}
\Delta \rho ({\bf r}) \approx  -{1 \over 2} \Biggl( {\partial
\Gamma({\bf r})
\over \partial \langle N \rangle} - {\Gamma({\bf r}) \over
\Delta^{2}(N)}  {\partial \Delta^{2}(N) \over \partial \langle N
\rangle}
\Biggr),
\end{equation} which inserted into (\ref{e11}) yields the following
approximation for $\Delta V_{\text{ext}}$:
 \begin{equation}
\beta\Delta V_{\text{ext}}({\bf r}) \approx {1 \over 2\Delta^{2}(N)}
{\partial
\Delta^{2}(N) \over \partial \langle N \rangle}  -{1 \over 2}
\int  d{\bf r}_{1} \, {\cal G}^{-1}({\bf r},{\bf r}_{1})  {\partial
\Gamma({\bf r}_{1}) \over \partial \langle N \rangle}
\label{e17}
\end{equation}   where we have used the identity
\begin{equation}
\label{e18}
\int  d{\bf r}_1\, {\cal G}^{-1}({\bf r},{\bf r}_1) \Gamma({\bf r}_1) =
1
\end{equation} which follows from Eqs. (\ref{e13}) and (\ref{e14})
[or, equivalently, from (\ref{b8}) and (\ref{b10})]. Considering the
derivative of Eq. (\ref{e18}) w.r.t. $\langle N \rangle$, and  using the
chain rule  and (\ref{e14}), Eq.  (\ref{e17}) can be re-expressed as
\begin{eqnarray}
\beta\Delta V_{\text{ext}}({\bf r}) &\approx& {1 \over
2\Delta^{2}(N)}
\Biggl( {\partial \Delta^{2}(N) \over \partial \langle N \rangle}  +
\int  d{\bf r}_{1}  d{\bf r}_{2}  {\delta {\cal G}^{-1}({\bf r},{\bf
r}_{1}) \over
\delta \rho_{\text{gc}}({\bf r}_{2}) }
\Gamma({\bf r}_{1}) \Gamma({\bf r}_{2})
\Biggr)\\ &=& {1 \over 2\Delta^{2}(N)}
\Biggl( {\partial \Delta^{2}(N) \over \partial \langle N \rangle} -
{\delta
\Delta^{2}(N) \over \delta \rho_{\text{gc}}({\bf r}) }
\Biggr)\, ,
\label{e19}
\end{eqnarray} where in the last equality we have used (\ref{b9b}).
Comparing (\ref{e19}) with (\ref{e5bisbisbis}) we obtain
\begin{equation}
\xi ({\bf r};[\rho_{\text{c}}])
\approx  -{1 \over 2\Delta^{2}(N)} {\delta \Delta^{2}(N) \over \delta
\rho_{\text{gc}}({\bf r}) } =\xi_{\text{sp}} ({\bf r};[\rho_{\text{gc}}])
\label{e21}
\end{equation} and
\begin{equation}
\xi (\rho_{0})
\approx  -{1 \over 2\Delta^{2}(N)} {\partial  \Delta^{2}(N) \over
\partial
\langle N \rangle} =\xi_{\text{sp}} (\rho_{0}).
\label{e21bis}
\end{equation} where $\xi_{\text{sp}}$ was defined in (\ref{b4}).
Finally, we note that in (\ref{e21}) it is shown that
$\xi[\rho_{\text{c}}]$ is approximately equal to
$\xi_{\text{sp}}[\rho_{\text{gc}}]$, i.e.,
$\xi_{\text{sp}}$ evaluated at $\rho_{\text{gc}}$ instead of
$\rho_{\text{c}}$. This approximate equality also holds for
$\xi_{\text{sp}}[\rho_{\text{c}}]$  as we show in what follows.
Expanding
$\xi_{\text{sp}}$ about  $\rho_{\text{gc}}$ we obtain
\begin{equation}
\xi_{\text{sp}} ({\bf r};[\rho_{\text{c}}]) =
\xi_{\text{sp}} ({\bf r};[\rho_{\text{gc}}]) +
\int d{\bf r}_1  {\delta\xi_{\text{sp}} ({\bf r};[\rho_{\text{gc}}])
\over \delta
\rho_{\text{gc}}({\bf r}_1) }
\Delta \rho + \cdots\,
\label{e23}
\end{equation} and, taking into account that $\xi$ is already a
$O(\Delta
\rho)$ quantity,
\begin{equation}
\xi_{\text{sp}} ({\bf r};[\rho_{\text{c}}])
\approx
\xi_{\text{sp}} ({\bf r};[\rho_{\text{gc}}]) + O(\Delta \rho)^2
\label{e23bis}
\end{equation} This proves the equivalence to first order of the two
approaches
 for obtaining an approximate density profile in the canonical
ensemble.

In summary, we have shown that two different approaches for
obtaining the density profile of a  fluid in the canonical ensemble are
equivalent to first order. The demonstration was based on
considering an external potential
${\bar V}_{\text{ext}}$ for which the equilibrium density in the grand
canonical ensemble is precisely the canonical ensemble result. Using
this external potential we have been able to work in the framework
of the grand canonical ensemble where approximations similar to
those carried out in this work are commonly encountered.

The proof of the equivalence gives additional support to the saddle
point approximation for the CE free-energy functional introduced in
\cite{White00}. This  approximation allows for a  CE-DFT treatment
of fluids confined in a closed cavity with excellent agreement with
simulation data. However, the SP free-energy functional was
proposed on the basis of the well-known SP relation between the
equilibrium CE free-energy and the grand potential of a homogeneous
fluid, assuming that this relation would be also a good approximation
for inhomogeneous fluids. This assumption is thus reinforced by  the
results of the present paper which are valid for any inhomogeneous
situation. As a final  remark, we would like to stress the fact that
our demonstration has focused on the approximate CE density rather
than on the free-energy functionals, and the equivalence between the
two approaches must be understood in this sense.

We  thank financial support by the Comisi\'on Interministerial de
Ciencia y Tecnolog\'{\i}a of Spain under Grant PB 98-0261.

\end{document}